\documentclass{aa}
\usepackage{graphicx}
\begin{document}

\title{The Mass of the Milky Way: Limits from a Newly Assembled Set of
Halo Objects}

\author{Tsuyoshi Sakamoto\inst{1} \and
        Masashi Chiba\inst{2} \and
        Timothy C. Beers\inst{3}}

\offprints{T. Sakamoto (sakamoto@pluto.mtk.nao.ac.jp)}

\institute{Department of Astronomical Science, The Graduate University
           for Advanced Studies, Mitaka, Tokyo 181-8588, Japan
          \and
           National Astronomical Observatory, Mitaka, Tokyo 181-8588, Japan
          \and
           Department of Physics \& Astronomy, Michigan State University, 
           East Lansing, MI 48824, USA}

\date{Received 7 August 2002; Accepted 14 October 2002}

\abstract{
We set new limits on the mass of the Milky Way, making use of the latest
kinematic information for Galactic satellites and halo objects.  Our sample
consists of 11 satellite galaxies, 137 globular clusters, and 413 field
horizontal-branch (FHB) stars up to distances of 10 kpc from the Sun.  Roughly
half of the objects in this sample have measured proper motions, permitting the
use of their full space motions in our analysis.  In order to bind these sample
objects to the Galaxy, their rest-frame velocities must be lower than their
escape velocities at their estimated distances.  This constraint enables us to
show that the mass estimate of the Galaxy is largely affected by several
high-velocity objects (Leo~I, Pal~3, Draco, and a few FHB stars), {\it not} by
a single object alone (such as Leo~I), as has often been the case in past
analyses. We also find that a gravitational potential that gives rise to a
declining rotation curve is insufficient to bind many of our sample objects to
the Galaxy; a possible lower limit on the mass of the Galaxy is about
$2.2\times 10^{12}~M_\odot$.  To be more quantitative, we adopt a Bayesian
likelihood approach to reproduce the observed distribution of the current
positions and motions of the sample, in a prescribed Galactic potential that
yields a flat rotation curve.  This method enables a search for the most likely
total mass of the Galaxy, without undue influence in the final result arising
from the presence or absence of Leo~I, provided that both radial velocities and
proper motions are used.  Although the best mass estimate depends somewhat on
the model assumptions, such as the unknown prior probabilities for the model
parameters, the resultant systematic change in the mass estimate is confined to
a relatively narrow range of a few times $10^{11} M_\odot$, owing to our
consideration of many FHB stars.  The most likely total mass derived from this
method is $2.5^{+0.5}_{-1.0}\times 10^{12}~M_{\odot}$ (including Leo~I), and
$1.8^{+0.4}_{-0.7}\times 10^{12}~M_{\odot}$ (excluding Leo~I).  The 
derived mass estimate of the Galaxy within the distance to the Large Magellanic
Cloud ($\sim 50$ kpc) is essentially independent of the model parameters,
yielding $5.5^{+0.0}_{-0.2}\times 10^{11}~M_{\odot}$ (including Leo~I) and
$5.4^{+0.1}_{-0.4}\times 10^{11}~M_{\odot}$ (excluding Leo~I).  Implications
for the origin of halo microlensing events (e.g., the possibility of brown
dwarfs as the origin of the microlensing events toward the LMC, may be excluded
by our lower mass limit) and prospects for more accurate estimates of the total
mass are also discussed.

\keywords{Galaxy: halo -- Galaxy: fundamental parameters --
 Galaxy: kinematics and dynamics -- stars: horizontal-branch}

}

\titlerunning{the Mass of the Milky Way}
\maketitle


\section{INTRODUCTION}

Over the past decades, various lines of evidence have revealed that the mass
density in the Milky Way is largely dominated by unseen dark matter, from
the solar neighborhood to the outer reaches of the halo (e.g.,
Fich \& Tremaine 1991).  Moreover, the presence of a dark component similar to
that found in our own Galaxy appears to be a generic feature in external
galaxies, as inferred from, e.g., flat rotation curves in their outer parts,
the presence of (a gravitationally bound) hot plasma in early-type galaxies,
and the observed gravitational lensing of background sources (e.g., Binney \&
Tremaine 1987).  A determination of the extent over which such
dark-matter-dominated mass distributions apply for most galaxies, including our
own, is of great importance for understanding the role of dark matter in galaxy
formation and dynamical evolution.  In particular, the mass estimate of the
Galaxy is closely relevant to understanding the origin of the microlensing
events toward the Large Magellanic Cloud (LMC) (e.g., Alcock et al. 2000;
Alcock et al. 2001).

While mass estimates of external galaxies can, in principle, be obtained in a
straightforward fashion using various dynamical probes, the total mass of the
Galaxy remains rather uncertain, primarily due to the lack of accurate
observational information for tracers located in its outer regions, where the
dark matter dominates.  The precise shape of the outer rotation curve, as
deduced from H~II regions and/or H~I gas clouds (e.g., Honma \& Sofue 1997), is
still uncertain because its determination requires knowledge of accurate
distances to these tracers (Fich \& Tremaine 1991).  Also, interstellar gas can
be traced only up to $\sim 20$ kpc from the Galactic Center, and hence provides
no information concerning the large amount of dark matter beyond this distance.

The most suitable tracers for determination of the mass distribution in the
outer halo of the Galaxy are the distant luminous objects, such as
satellite galaxies, globular clusters, and halo stars on orbits that explore
its farthest reaches (e.g., Miyamoto et al. 1980; Little \&
Tremaine 1987; Zaritsky et al. 1989; Kochanek 1996; Wilkinson \& Evans 1999,
hereafter WE99).  However, the limited amount of data presently available
on the full space motions of these tracers, and the small size of the available
samples, have stymied their use for an accurate determination of the Galaxy's
mass. In particular, most previous mass estimates (except for WE99, see below)
depend quite sensitively on whether or not a distant satellite, Leo~I, is bound
to the Galaxy.  Leo~I has one of the largest radial velocities of the known
satellites, despite its being the second most distant satellite from the Galaxy
(Mateo 1998; Held et al. 2001).  As a consequence, estimates of the total mass
of the Galaxy are much more uncertain (by as much as an order of magnitude)
than, for instance, the value of the circular speed in the solar neighborhood
(Kerr \& Lynden-Bell 1986; Fich \& Tremaine 1991; Miyamoto \& Zhu 1998;
M\'{e}ndez et al. 1999).

Recently, by making use of both the observed radial velocities and proper
motions of six distant objects, WE99 demonstrated that the use of full space
motions can provide a reliable mass estimate of the Galaxy without being
largely affected by the presence or absence of Leo~I. They also argued that the
primary uncertainties in their mass estimate arose from the small size of the
data set and the measurement errors in the full space motions, especially the
proper motions.  This work motivated us to investigate a much larger data set,
with more accurate kinematic information, to set tighter limits on the mass of
the Galaxy.  Specifically, as we show below, there are two objects among the
WE99 sample (Draco and Pal~3) that have relatively large velocity errors, yet
still play crucial roles in a determination of the Galaxy's mass, so the
addition of more (and better) data is important.

Over the past few years, the number of distant satellite galaxies and globular
clusters with available proper motions has gradually increased (e.g., Mateo
1998; Dinescu et al. 1999; Dinescu et al. 2000; Dinescu et al.
2001).  In addition, another tracer population that is suitable for exploring
mass estimates of the Galaxy has become available from the extensive
compilation of A-type metal-poor stars by Wilhelm et al. (1999b), which
provided radial velocity measurements, as well as estimates of the physical
parameters of these stars (e.g., [Fe/H], T$_{\rm eff}$, log $g$).  
Among the Wilhelm et al.  sample, the luminous FHB stars are the most
useful mass tracers, both because of their intrinsic brightness, and the fact
that accurate distance determinations can be inferred from their absolute
magnitudes on the horizontal branch (e.g., Carretta et al.  2000). Moreover,
there exist proper-motion measurements for many of these stars, provided by
both ground- and space-based proper-motion catalogs (Klemola et al. 
1994; R\"oser 1996; Platais et al. 1998; Hog et al.  2000), from which full
space motions may be derived.

In this paper we re-visit the mass determination of the Galaxy, based on a
sample of 11 satellite galaxies, 137 globular clusters, and 413 FHB stars, out
of which 5 satellite galaxies, 41 globular clusters, and  211 FHB stars have
measured proper motions.  First, we investigate the lowest possible mass that
the Galaxy may have, adopting the requirement that the rest-frame velocities of
observed sample objects be less than their escape velocities at their present
distance from the Galactic center (e.g., Miyamoto et al. 1980; Carney et al. 
1988). Secondly, the most likely mass of the Galaxy is calculated,
based on a Bayesian likelihood analysis that seeks to reproduce both the
current positions and velocities of the sample objects (e.g., Little \&
Tremaine 1987; Kochanek 1996; WE99).  Because our sample of tracers is, by far,
the largest and most accurate one presently available, it is possible to place
more reliable limits on the total mass of the Galaxy.  In Sect. 2 we describe
our sample objects and the assembly of their kinematic data. In Sect. 3 
we discuss the influence of the adopted membership of our sample on
the mass determination.  In Sect. 4 we obtain the most likely total
mass of the Galaxy based on a Bayesian likelihood analysis.  In Sect. 5 we
discuss implications for the origin of the halo microlensing events toward the
LMC and the mass estimate of the Local Group, and consider the prospects for
more obtaining more accurate estimates of the total mass of the Galaxy in the
near future.


\section{DATA}

We consider a sample of objects that serve as tracers of the Galactic mass
distribution consisting of 11 satellite galaxies, 137 globular clusters, and
413 FHB stars.  In the case of the satellite galaxies, all of the basic
information for our kinematic analysis, i.e., positions, heliocentric
distances, and heliocentric radial velocities, are taken from the compilation
of Mateo (1998).  For the globular clusters, we adopt the information provided
by Harris (1996), including their positions and heliocentric radial velocities,
their metal abundances, [Fe/H], and the apparent magnitude of the clusters'
horizontal branch (HB).  The catalog of Wilhelm et al. (1999b) is our source of
similar information for the FHB stars.  We obtain an internally consistent set
of distance estimates for the globular clusters and the FHB stars from the
recently derived relationship between the absolute magnitude of the HB,
$M_V$(HB), and [Fe/H], by Carretta et al. (2000),

\begin{equation}
M_V {\rm (HB)} = (0.18\pm0.09) ([{\rm Fe/ H}]+1.5)+ (0.63\pm0.07) \ .
\end{equation}

\noindent Clearly, we have assumed that there is no large offset between the
absolute magnitudes of FHB stars and their counterpart HB stars in the globular
clusters (a view also supported by the recent work of Carretta et al. 
2000).  Fig. 1 shows the spatial distribution of the globular
clusters, satellite galaxies, and FHB stars on the plane perpendicular to the
Galactic disk, where $X$ axis connects the Galactic center ($X$=0) and the Sun
($X$=8.0 kpc).  The filled and open symbols denote the objects with and without
proper-motion measurements, respectively.  Satellite galaxies are the most
distant tracers, with Galactocentric distances, $r$, greater than 50 kpc.  The
globular clusters extend out to almost $r = 40$ kpc, while the present sample
of FHB stars are confined to locations within 10 kpc of the Sun.  Thus, our
sample objects are widely, though not uniformly, distributed throughout the
volume of the Galaxy.

\begin{figure}
\resizebox{\hsize}{!}{\includegraphics{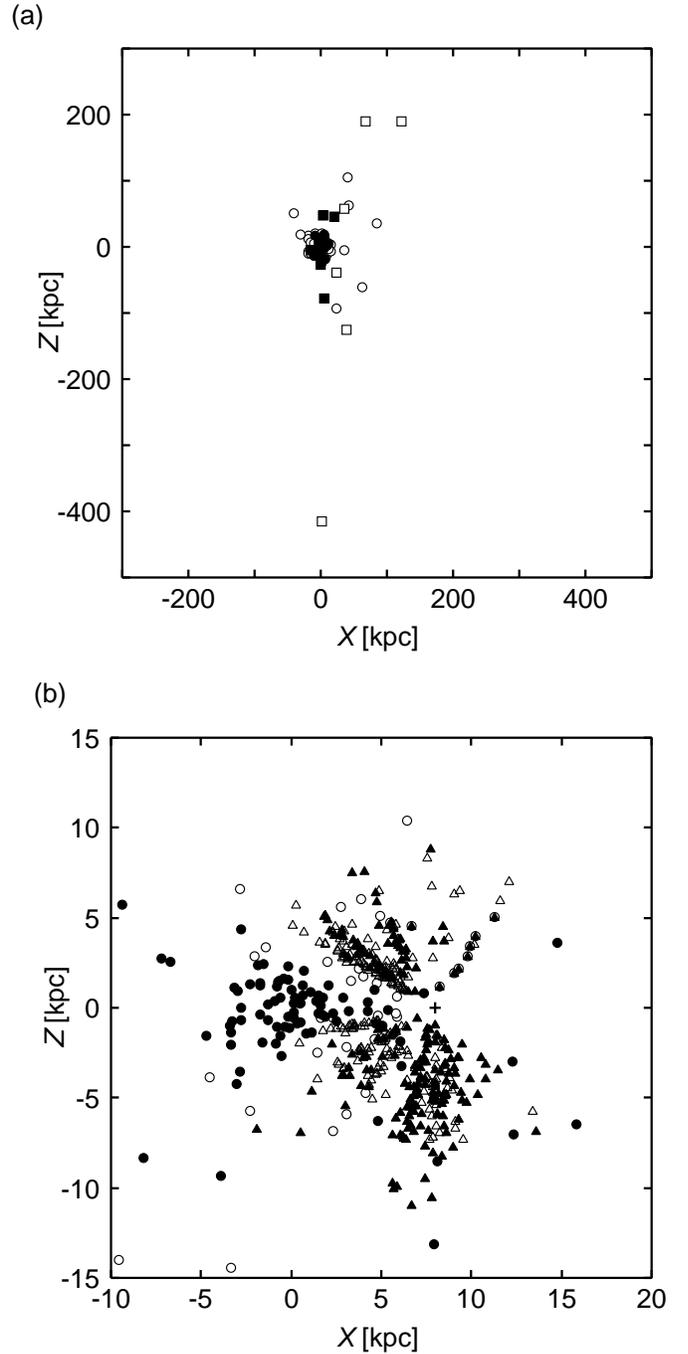}}

\caption{
Spatial distributions of satellite galaxies (squares), globular clusters
(circles), and FHB stars (triangles) on the plane perpendicular to the Galactic
disk, where the $X$ axis connects the Galactic center ($X$ =0) and the Sun
($X$=8.0 kpc).  The filled and open symbols denote the objects with and without
available proper motions, respectively.  The plus sign in panel (b) denotes the
position of the Sun, $(X, Y)=(8.0, 0)$.
} 
\end{figure}

Among these sample objects there exist proper-motion measurements for 5 of the
satellite galaxies, 41 of the globular clusters, and 211 of the FHB stars.  The
proper motion data for the LMC, Sculptor, and Ursa Minor are taken from WE99,
whereas those for Sagittarius and Draco are taken from Irwin et al. (1996) and
Scholz \& Irwin (1994), respectively.  The proper motions for most of the
globular clusters have been compiled by Dinescu et al. (1999). We adopt the
data from this source, except for two globular clusters with recently revised
proper-motion measurements (NGC~6254: Chen et al. 2000; NGC~4147: Wang et al.
2000), and for three additional globular clusters compiled recently (Pal~13:
Siegel et al.  2000; Pal~12: Dinescu et al.  2000; NGC~7006: Dinescu et al.
2001).  Proper motions for 211 of the FHB stars in the Wilhelm et al.  (1999b)
sample are available from one or more existing proper-motion catalogs.  These
include the STARNET Catalog (R\"oser 1996), the Yale-San Juan Southern Proper
Motion Catalog (SPM~2.0: Platais et al. 1998), the Lick Northern Proper Motion
Catalog (NPM1: Klemola et al. 1994), and the TYCHO-2 Catalog (Hog
et al.  2000).  Many of these FHB stars have been independently measured in two
or more catalogs, so that by combining all measurements one can reduce the
statistical errors, as well as minimize any small remaining systematic errors
in the individual catalogs, as was argued in Martin \& Morrison (1998) and
Beers et al. (2000).

We estimate average proper motions, $<\mu>$, and their errors,
$<\sigma_{\mu}>$, weighted by the inverse variances, as

\begin{eqnarray}
<\mu>&=&(\sum_{i=1}^n\mu_i/\sigma_{\mu_i}^2)
/(\sum_{i=1}^n1/\sigma_{\mu_i}^2),\\
<\sigma_{\mu}>&=&(\sum_{i=1}^n 1/\sigma_{\mu_i}^2)^{-1/2},
\end{eqnarray}

\noindent where $n$ denotes the number of measurements.  
Table 1 (available at the CDS) lists these
compilations, as well as the estimated distances to the FHB stars, where $r$
and RV denote the Galactocentric distances and heliocentric radial velocities,
respectively.  Typical errors in the reported proper-motion measurements range
from 1 to $ \sim 5$ mas~yr$^{-1}$ for individual field stars, whereas those for
the satellite galaxies and globular clusters are about 0.3 mas~yr$^{-1}$ and 1
mas~yr$^{-1}$, respectively.

We assume a circular rotation speed for the Galaxy of $V_{\rm LSR}=220$ km~${\rm
s}^{-1}$ at the location of the Sun (i.e. $R_{\odot}=8.0$ kpc along the disk
plane) and a solar motion of ($U, V, W$)= ($-$9, 12, 7) km~${\rm s}^{-1}$
(Mihalas \& Binney 1981), where $U$ is directed outward from the Galactic
Center, $V$ is positive in the direction of Galactic rotation, and $W$ is
positive toward the North Galactic Pole\footnote{ Dehnen \& Binney (1998)
derived the solar motion of $(U, V, W) = (-10.0, 5.3, 7.2)$ km~s$^{-1}$
based on Hipparcos data. Because the difference between this and the currently
adopted solar motion is only a few km~s$^{-1}$, it gives little influence on
the mass estimate.}.  We then calculate the space motions
of our tracers, as well as the errors on these motions, fully taking into
account the reported measurement errors in the radial velocities of the
individual satellite galaxies (typically a few km~s$^{-1}$), adopting a typical
radial-velocity error for other objects (10 km~s$^{-1}$), the measurement
errors assigned to the proper motions of each object (when available, adopting
a mean error for the source catalog when not), and distance errors for the
satellite galaxies (10 \% relative to the measured ones), or as obtained from
Eq. 1 for the globular clusters and FHB stars.

It is worth noting that the reported proper motions of the FHB stars in our
sample may yet contain unknown systematics with respect to their absolute
motions in a proper reference frame; this caution applies to the globular
clusters and satellite galaxies as well. It is an important goal to make
efforts to reduce the systematic, as well as random, errors in the proper
motions upon which studies of Galactic structure and kinematic studies are
based, using much higher precision astrometric observations than have been
obtained to date.


\section{ROLES OF THE SAMPLE IN THE MASS LIMITS}

If we model the Galaxy as an isolated, stationary mass distribution, and assume
that all of our tracer objects are gravitationally bound to it, then the
rest-frame velocities of all objects, $V_{\rm RF}$, must be less than their escape
velocities, $V_{\rm esc}=\sqrt{2\psi}$, where $\psi$ denotes the gravitational
potential of the Galaxy. A number of previous researchers have adopted this
method for estimation of the mass of the Galaxy (e.g., Fricke 1949; Miyamoto
et al. 1980; Carney et al. 1988; Leonard \& Tremaine 1990; Dauphole \& Colin
1995).  Note that the mass estimate obtained in this fashion is largely
dominated by only a small number of high-velocity objects, hence the mass
that is derived depends rather sensitively on the selection criteria
adopted for such objects.  In the present section, instead of deriving an exact
mass estimate, we follow this procedure for the purpose of elucidating the role
of sample selection in setting limits on the mass of the Galaxy.

To this end we consider two different mass models, in order to investigate
the difference in the mass limits obtained by the use of different potentials.
Our models, hereafter referred to as Model A and B, are the same as those
adopted in WE99 and Johnston et al. (1995) (and also used by
Dinescu et al. 1999), respectively.

Model A has spherical symmetry, and results in a flat rotation curve in the
inner regions of the Galaxy. The gravitational potential and mass density are
given as:

\begin{eqnarray}
\psi(r) &=& \frac{GM}{a} \log \left(
\frac{\sqrt{r^2+a^2}+a}{r} \right) \ , \nonumber\\
\rho(r) &=& \frac{M}{4\pi}\frac{a^2}{r^2(r^2+a^2)^{3/2}} \ ,
\end{eqnarray}

\noindent where $a$ is the scale length of the mass distribution, and $M$ is
the total mass of the system. The central density of this model is cusped (like
$r^{-2}$) and falls off as $r^{-5}$ for $r \gg a$. 
 Using Eq. (4), the circular rotation speed is given as 
$V_{\rm c}^2 = GM/(r^2+a^2)^{1/2}$, so by setting $V_{\rm c}$ at $r=R_\odot$ as $V_{\rm LSR}=$
220 km~s$^{-1}$ in our standard case, it follows that this model contains
one free parameter, $a$, to obtain $M$.

\begin{figure}
\resizebox{\hsize}{!}{\includegraphics{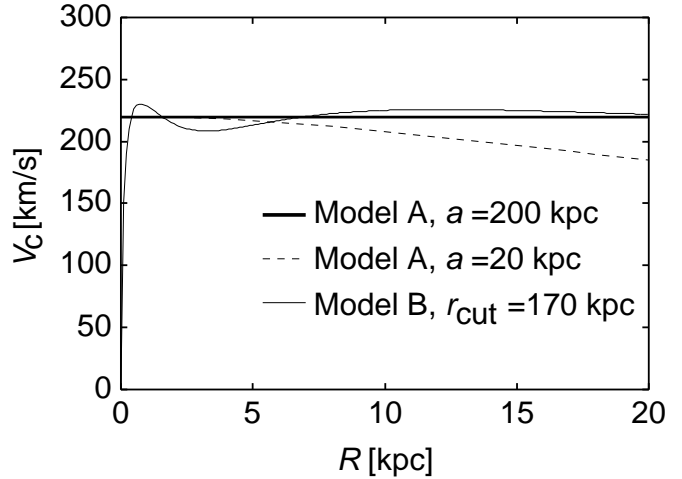}}
\caption{
Rotation curves for Model A and Model B, parameterizations of
the mass distributions considered in this paper.  See the text for more
information on the nature of these models.
} 
\end{figure}

Model B consists of more realistic axisymmetric potentials with three
components (the bulge, disk, and dark halo) that reproduce the shape of the
Galactic rotation curve (Johnston et al. 1995). The bulge and disk components
we adopt are those represented by Hernquist (1990) and Miyamoto \& Nagai (1975)
potentials, respectively. All of the parameters included in these potentials
are taken from Dinescu et al.  (1999) (see their Table 3).  In order to obtain
a finite total mass we assume the following modified logarithmic potential
(corresponding to an isothermal-like density distribution) for the dark halo
component:

\begin{equation}
\psi_{\rm halo}(r) = 
\left \{
\begin{array}{l}
v_0^2 \log [1+(r/d)^2]-\psi_0 , \qquad {\rm at}~r<r_{\rm cut} \\
-2v_0^2 \frac{r_{\rm cut}}{r}\frac{c}{1+c} , \qquad {\rm at}~r \ge r_{\rm cut} \ ,
\end{array}
\right. \
\end{equation}
\begin{equation}
\rho(r)=\frac{2v_0^2}{4\pi Gd^2}\frac{3+r/d}{(1+r/d)^3} ,
\end{equation}
where $\psi_0$ is defined as
\begin{equation}
\psi_0=v_0^2 [ \log (1+c)+ 2c/(1+c) ] , \quad c=(r_{\rm cut}/d)^2 \ ,
\end{equation}

\noindent and we adopt $v_0 = 128$ km~s$^{-1}$ and $d = 12$ kpc (Dinescu et al.
1999).  This model contains one free parameter, namely the cutoff radius of the
dark halo, $r_{\rm cut}$.  Fig. 2 shows the rotation curves for $0 \le R \le
20$ kpc, provided by Model A with $a=200$ kpc (thick solid line) and Model B
with $r_{\rm cut}=170$ kpc (thin solid line), where both curves shown at $R
\le 20$ kpc remain unchanged as long as $a, r_{\rm cut} \gg 20$ kpc.  The circular
speed at $R=R_\odot$ is 220 $\rm{km~s^{-1}}$ for both mass models.
Also shown is the declining rotation curve with increasing radius, as obtained
from Model A with $a=20$ kpc and $V_{\rm LSR}=211$ km~s$^{-1}$ (dashed line).

\begin{figure}
\resizebox{\hsize}{!}{\includegraphics{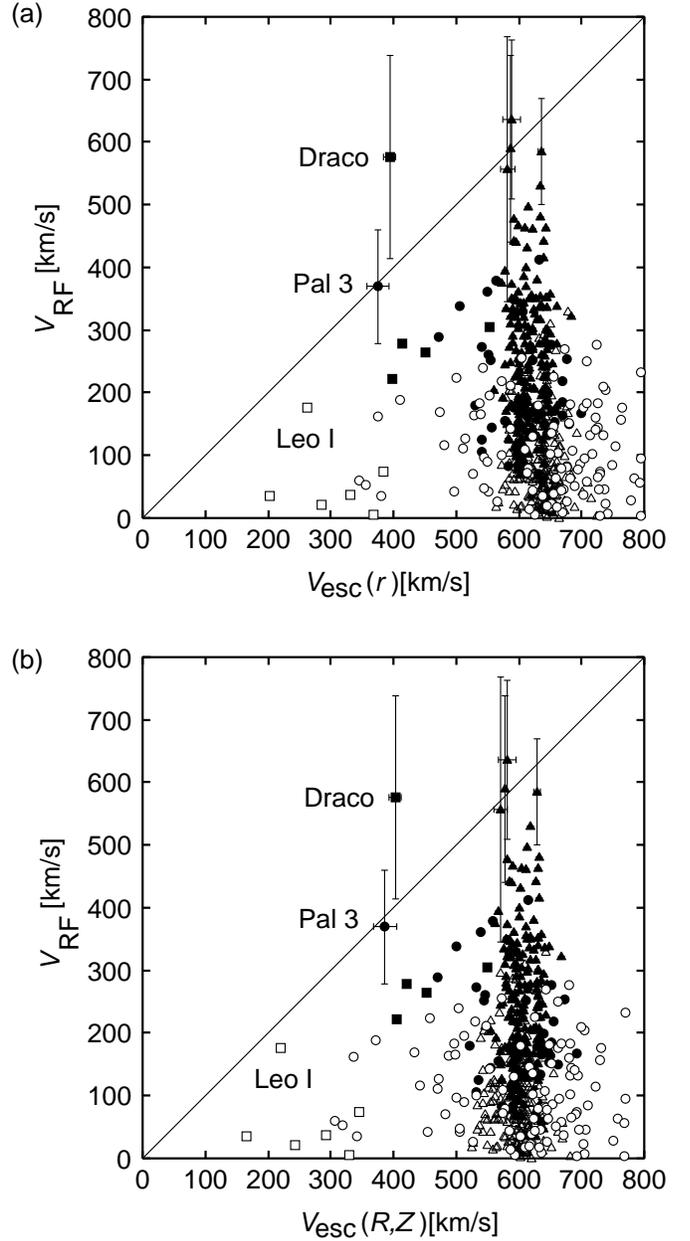}}
\caption{
(a) The relation between escape velocities, $V_{\rm esc}$, and 
space velocities, 
$V_{\rm RF}$, for Model A with $a=195$ kpc and $V_{\rm LSR}=220$ km~s$^{-1}$.
The symbols are the same as those in
Fig. 1.  The solid line denotes the boundary between the gravitationally
bound and unbound objects -- those in the region below the line are bound
to the Galaxy. For the sake of clarity, velocity errors are plotted for only
the high-velocity objects relevant to the mass estimate.  (b) Same as panel (a)
but for Model B with $r_{\rm cut}=295$ kpc.
} 
\end{figure}

We plot, in Fig. 3a and 3b, the relationship between the derived escape
velocities, $V_{\rm esc}$, and the rest-frame velocities, $V_{\rm RF}$, when we 
adopt 
Model A with $a=195$ kpc and Model B with $r_{\rm cut}=295$ kpc, respectively.
For the objects without available proper motions (open symbols), we adopt the
radial velocities alone as measures of $V_{\rm RF}$, hence their estimated space
velocities are only lower limits.  The solid line denotes the boundary between
the objects that are bound (below the line) and unbound (above the line) to
the Galaxy, respectively, where the positions of the data points relative
to the solid line depend on the choice of $a$ or $r_{\rm cut}$. As these figures
clearly indicate, the selection of the smallest value of $a$ or
$r_{\rm cut}$ that places the sample inside the bound region, or equivalently
the lower limit on the mass of the Galaxy, is controlled by
a few high-velocity objects located near the boundary line at each respective
radius (or corresponding $\psi$).  These objects include Leo~I (for which only
radial velocity information is available), Draco, Pal~3, and several
high-velocity FHB stars. This highlights the following important properties of
the derived mass limits: (1) If the proper motions of all objects are
unavailable, then the mass estimate sensitively depends on the presence or
absence of Leo~I, as has been noted in previous studies. (2) Compared to case
(1), if the available proper motions of the satellite galaxies and globular
clusters are taken into account, the constraint provided by Draco and Pal~3 is
basically the same as that provided by Leo~I.  This may explain the result of
WE99, which showed that the mass determination is made insensitive to Leo~I if
the proper motion data of satellite galaxies and globular clusters are taken
into account.  However, as Fig. 3 indicates, the velocity errors for Draco
and Pal~3 are quite large, so these objects place only weak constraints on the
mass estimate.  (3) If we consider the proper motions of FHB stars, then some
FHB stars having high velocities provide the basically the same constraint on
the Galaxy's mass as do Leo~I, Draco, and Pal~3.  These properties suggest that
the inclusion of FHB stars with available proper motions is crucial, and that
they provide constraints on the mass limit of the Galaxy that depend on {\it
neither} the inclusion or absence of Leo~I {\it nor} on the large velocity
errors for Draco and Pal~3.

As mentioned above, a determination of the lower mass limit for the Galaxy,
using escape velocities, inevitably depends on the selection of a few {\it
apparently} high-velocity objects from a much larger sample of tracers. While
a mass estimate independent of this selection effect will be obtained in
Sect. 4, we seek first to obtain a rough measure for the lower mass
limit, i.e., the smallest $a$ or $r_{\rm cut}$ that encloses Pal~3, Draco, and the
four highest-velocity FHB stars (for which we mark asterisks after their names
in Table 1) inside the bound region, based on a weighted least-squares fitting
procedure (weights being inversely proportional to the velocity errors).  This
exercise yields $a = 195^{+160}_{-85}$ kpc for Model A, and
$r_{\rm cut}=295^{+335}_{-145}$ kpc for Model B. Using these values, the lower
limits to the total mass, $M$, of the Galaxy may be given as $2.2^{+1.8}_{-1.0}
\times 10^{12} M_{\odot}$ for Model A and $2.2^{+2.6}_{-1.1}
\times 10^{12} M_{\odot}$ for Model B, respectively.  Thus, the difference in
the derived mass limits is not significant, as long as the rotation curve at
outer radii is approximately constant at the adopted value of 220 km~s$^{-1}$.
It also suggests that the flattened nature of the Model B potential, due to the
presence of the disk component, does not affect the results significantly --
the high-velocity tracers are located at large Galactocentric distances and/or
their orbits largely deviate from the disk plane.

In addition to the above experiments, we also consider a mass model that
yields a declining rotation curve at outer radii, as was proposed by Honma \&
Sofue (1997) from their H~I observations. We adopt Model A with $a=20$ kpc
and $V_{\rm LSR} = 211$ km~s$^{-1}$ at $R=R_\odot$, so as to yield the declining
rotation curve at $R > R_\odot$ (dashed line in Fig. 2), which is
reminiscent of the result in Honma \& Sofue (1997). Fig. 4 shows the
$V_{\rm RF}$ vs. $V_{\rm esc}$ relationship that follows from adoption of this model.
As is evident, the total mass obtained from a model that leads to a declining
rotation curve is insufficient to bind many of our sample objects to the
Galaxy.

\begin{figure}
\resizebox{\hsize}{!}{\includegraphics{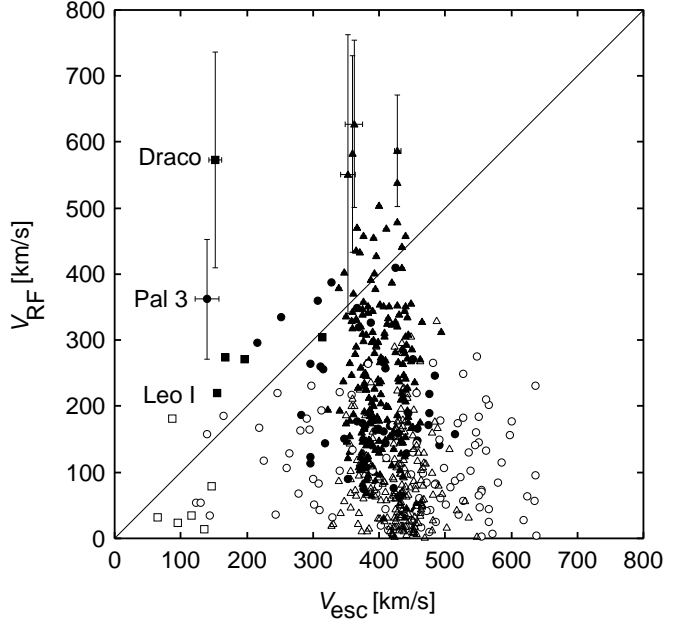}}
\caption{
The relation between escape velocities, $V_{\rm esc}$, and space
velocities, $V_{\rm RF}$, for Model A with $a=20$ kpc and
$V_{\rm LSR}=211$ km~s$^{-1}$. In this case, the rotation
curve declines with increasing radii, as shown in Fig. 2 (dashed line). Note
that, if this situation were to apply,  many of the sample objects would be
unbound to the Galaxy.
} 
\end{figure}

\section{MASS DETERMINATION BASED ON A BAYESIAN LIKELIHOOD METHOD}

\subsection{Method}

To obtain a more quantitative measure of the mass of the Galaxy, we examine
an alternative method that takes into account {\it all of the positional and
kinematic information} of the sample objects, in contrast to the use of the
high-velocity tracers alone, as in the previous section.  In this approach, a
phase-space distribution function of tracers, $F$, is prescribed for a
specifically chosen $\psi$, and the model parameters included in $F$ and $\psi$
are derived so as to reproduce the presently observed positions and velocities
of the tracers in the (statistically) most significant manner. The optimal
deduced parameters relevant to $\psi$ then allow us to estimate the total mass
of the Galaxy. This method was originally proposed by Little \& Tremaine
(1987), and further developed by Kochanek (1996) and WE99.

Based on the results presented in the previous section, we take Model A with
spherical symmetry as the mass distribution of the Galaxy, which is sufficient
for the following analysis.  For the sake of simplicity, and also for ease of
comparison with the previous studies by Kochanek (1996) and WE99, the
phase-space distribution function is taken to have the same anisotropic form as
that adopted in these studies.  That is, it depends on the binding energy per
unit mass, $\varepsilon$ ($\equiv \psi - v^2/2$), and the angular momentum per
unit mass, $l$, in the following way:

\begin{equation}
F(\varepsilon,l)=l^{-2\beta}f(\varepsilon) \ ,
\end{equation}
where 
\begin{eqnarray}
f(\varepsilon)&=&
\frac{2^{\beta -3/2}}{\pi^{3/2}\Gamma [m-1/2+\beta] \Gamma[1-\beta]}
\frac{d}{d\varepsilon}\nonumber\\
&\times&\int_{0}^{\varepsilon}d\psi \frac{d^mr^{2\beta}\rho_{\rm s}}
{d\psi^m}(\varepsilon-\psi)^{\beta-3/2+m} \ ,
\end{eqnarray}

\noindent where $\rho_{\rm s}$ is the tracer density distribution, $\Gamma$ is the
gamma function, and $m$ is an integer whose value is chosen such that the
integral in Eq. (9) converges (e.g., Dejonghe 1986; Kochanek 1996). In the
spherical model, this form of the distribution function yields equal velocity
dispersions in the orthogonal angular directions, $<v_\theta^2> = <v_\phi^2>$,
and a constant anisotropy $\beta=1-<v_{\theta}^2>/<v_r^2>$ everywhere in the
Galaxy.  Our choice of $m=2$ in Eq. (9) (to be in accord with the WE99
work) limits the allowed range for the velocity anisotropy to $-1.5
\leq\beta\leq 1$ when proper motion data are considered, while the use of
radial velocities alone sets no limit for tangential anisotropy [$-\infty, 1$].

For $\rho_{\rm s}$, we consider WE99's two models: (a) Shadow tracers following the
mass density distribution obtained from Model A (Eq. 4), and (b) a power-law
distribution as a function of $r$.  The shadow-tracer model is given as:

\begin{equation}
\rho_{\rm s}(r)\propto\frac{a_s^2}{r^2(r^2+a_s^2)^{3/2}} \ ,
\end{equation}

\noindent where $a_s$ is the scale length.  The power-law model with
index $\gamma$ is given as:

\begin{equation}
\rho_{\rm s}(r)\propto\frac{1}{r^{\gamma}} \ .
\end{equation}

\noindent It should be noted that, since shadow tracers may be truncated at the
distance below the scale length of the mass distribution, the scale length of
the tracers, $a_s$, is generally different from the scale length of the
Galaxy's mass, $a$.

\begin{figure}
\resizebox{\hsize}{!}{\includegraphics{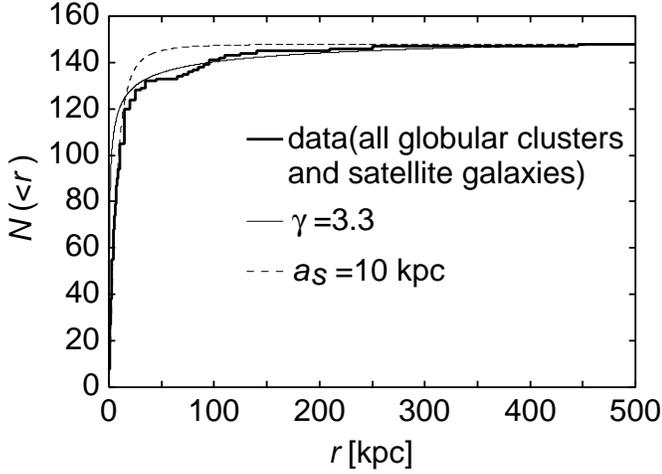}}
\caption{
Cumulative number distribution, $N(<r)$, of the distances of
globular clusters and satellite galaxies (solid histogram) in comparison with
model distributions (continuous dashed and solid lines).  See the text for
additional information.
} 
\end{figure}

Using the 27 objects (satellite galaxies and globular clusters) at $r > 20$
kpc, WE99 derived $a_s = 100$ kpc and $\gamma = 3.4$ as the best fitting
parameters for their spatial distribution. We re-examine $a_s$ and $\gamma$
using our sample of all satellite galaxies and globular clusters.  Note that
the FHB stars are excluded in this determination of $a_s$ and $\gamma$, as they
have not (yet) been completely surveyed over the Galactic volume.  We obtain
$a_s = 10$ kpc and $\gamma = 3.3$ as the best-fit values, based on a simple
K-S test of the observed vs. predicted distribution functions (see Fig. 5).
If we exclude the globular clusters at $r \le 10$ kpc, for which the spherical
symmetry assumption may be questionable due to the presence of the disk
globular clusters, we obtain $a_s = 50$ kpc and $\gamma = 3.4$.  Thus, $a_s$
depends sensitively on the adopted range of radius (i.e., on the
selection of the sample), whereas $\gamma$ basically remains unchanged.
Therefore, we focus our attention on the results obtained from the power-law
representation for the tracer population; the shadow-tracer population is
also examined for the purpose of comparison with WE99.  To see the dependence
of the mass estimate on these parameters, we obtain estimates for two values of
$\gamma$ (3.4 and 4.0) and $a_s$ (100 kpc and the scale length of the mass
distribution, $a$), respectively. We note that the FHB stars are also expected
to follow a power-law form with $\gamma \simeq 3.4$, as inferred from other
halo field stars (e.g., Preston et al. 1991; Chiba \& Beers 2001).

We calculate the likelihood of a particular set of model parameters
(the scale length of the mass distribution, $a$, and the anisotropy parameter,
$\beta$) given the positions, $r_i$, and radial velocities, $v_{ri}$, or space
velocities, $v_{i}$, using Bayes' theorem.  The probability that the model
parameters take the values $a$ and $\beta$, given the data $(r_i,v_{(r)i})$ and
prior information $I$, is:

\begin{equation}
P(a,\beta |r_i,v_{(r)i},I)=\frac{1}{N}P(a)P(\beta)\prod_{i=1}^{N}
P(r_i,v_{(r)i}|a,\beta) \ ,
\end{equation}

\noindent where $N$ is the normalization factor (Kochanek 1996; WE99).  The
probabilities $P(a)$ and $P(\beta)$ denote the prior probability distributions
in $a$ and $\beta$, respectively.  Here, $P(r_i,v_{(r)i}|a,\beta)$ corresponds
to the probability of finding an object at position $r_i$ moving with radial
velocity $v_{(r)i}$, or space velocity $v_{i}$, for a particular set of model
parameters $a$ and $\beta$.  The complete expressions for
$P(r_i,v_{(r)i}|a,\beta)$ are shown in Table 1 of WE99. To calculate this
probability for the objects with full space velocities, we take into account
their large errors relative to radial velocities alone (due to the observed
proper-motion and assumed distance errors), by multiplying by an error
convolution function of the form:

\begin{eqnarray}
P(r_i,v_i|a,\beta) &=& \int\int dv_{\alpha}dv_{\delta}
E_1(v_{\alpha})E_1(v_{\delta}) \nonumber\\
&\times& P(r_i,v_{i,obs}(v_{\alpha},v_{\delta})|a,\beta) \ ,
\end{eqnarray}

\noindent where $(v_{\alpha},v_{\delta})$ are the tangential velocities along
the right ascension and declination coordinates, respectively, and $E_1$ is the
Lorentzian error convolution function, defined as:

\begin{equation}
E_1(v)=\frac{1}{\sqrt{2}\pi\sigma_1}\frac{2\sigma_1^2}{2\sigma_1^2+
(v-v_{obs})^2} \ ,
\end{equation}

\noindent where $\sigma_1$ is defined as $\sigma_1=0.477\sigma$ for the
calibrated error estimate $\sigma$ (see WE99).

The prior probability in the velocity anisotropy, $\beta$, is taken to be of
the form $P(\beta) \propto 1/(3-2\beta)^n$, where $n=0$ and 2 correspond to a
uniform prior and a uniform energy prior, respectively (Kochanek 1996; WE99).
Larger values of $n$ yield a larger weight towards radial anisotropy.  For the
prior probability in $a$, $P(a)$, we adopt $1/a$ and $1/a^2$ (WE99).

Using the routine AMOEBA in Numerical Recipes (Press et al. 1992), we search
for a set of model parameters, $a$ and $\beta$, that maximize the probability
$P(a,\beta |r_i,v_{(r)i},I)$. The total mass of the Galaxy, $M$, is then
derived from the parameter $a$.
 
\subsection{Results}

Initially, we apply the Bayesian likelihood method, making use of only the
radial velocities of the objects, setting aside for the moment the available
proper-motion information.  Specifically, we focus on the difference in the
mass estimate arising from the presence or absence of Leo~I.  Fig. 6 shows
the likelihood contours in the mass-anisotropy ($M-\beta$) plane for the case
of a power-law tracer population with $\gamma=3.4$, where $\beta$ is limited to
the range of $-1.5 \le \beta \le 1$. The solid and dashed lines denote the
presence and absence of Leo~I, respectively. As is evident, the mass estimate
sensitively depends on whether or not Leo~I is bound to the Galaxy, as has been
noted in previous studies. Inclusion of Leo~I yields a likely total mass that
is {\it an order of magnitude greater} than the case without Leo~I.  Over the
range of $\beta$ we consider, the most likely value of $M$, with Leo~I included
,is $21.0\times 10^{11} M_\odot$, corresponding to a scale length $a = 185$
kpc, whereas excluding Leo~I yields $M = 9.6\times 10^{11} M_\odot$, and $a =
85$ kpc. We note that the role of Leo~I in the Galaxy's mass estimate is also
understandable from the escape-velocity argument;  if only the sample radial
velocities are taken into account, Leo~I {\it alone} determines the best-fit
boundary line $V_{\rm RF}=V_{\rm esc}$ in the $V_{\rm RF}$ vs. $V_{\rm esc}$ diagram (Fig. 3).

\begin{figure}
\resizebox{\hsize}{!}{\includegraphics{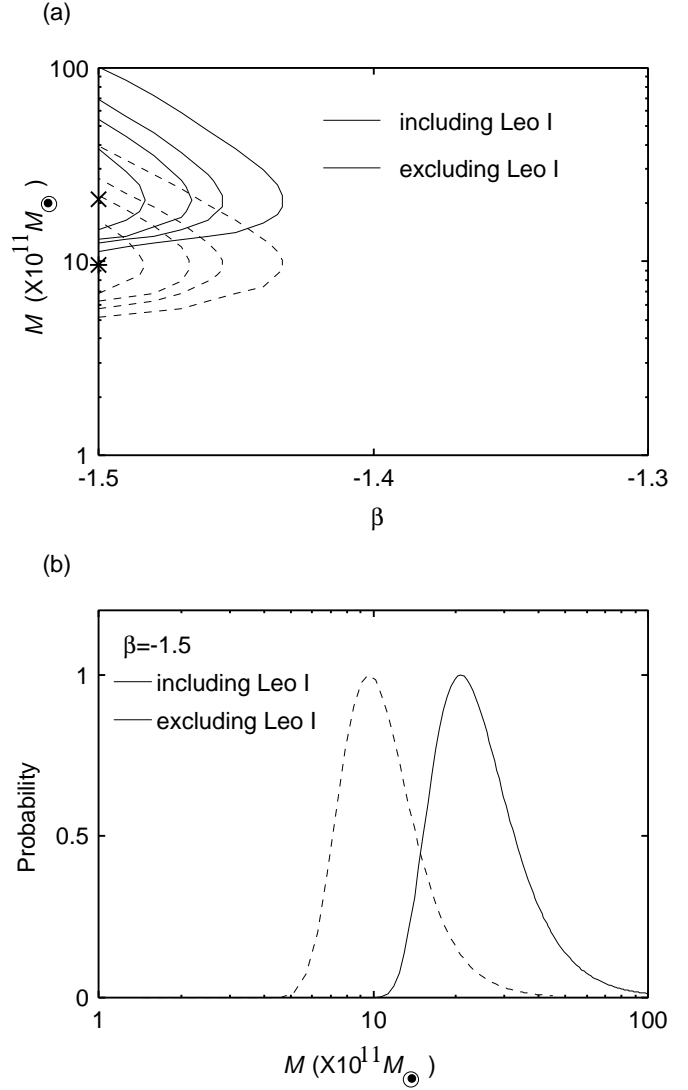}}
\caption{
(a) Likelihood contours in the plane of the mass, $M$, and velocity
anisotropy, $\beta$, obtained from an analysis using only radial velocities.
The solid and dashed curves show the results including Leo I and excluding Leo
I, respectively; the cross and the asterisk show the maxima of the
probabilities for each case. Contours are plotted at heights of 0.32, 0.1,
0.045, and 0.01 of the peak height. The spatial distribution of a tracer
population is assumed to follow a power-law form with $\gamma=3.4$.  (b)
Probabilities of the mass $M$ at $\beta=-1.5$, including Leo~I (solid line) and
excluding Leo~I (dashed line).
}
\end{figure}

As is seen in Fig. 6, the high-probability region is biased toward the line
$\beta =-1.5$. This bias arises from the specific form of the phase-space
distribution function, $F(\varepsilon,l)$, given in Eq. (8), where the
probability $P(a, \beta| r_i, v_{(r)i}, I)$ is high at large $F$. We plot $F$
in Fig. 7 for a set of $r$ and $\beta$ (solid and dotted lines for $\beta=-1$
and 1, respectively).  It follows that $F$ at high $\varepsilon$ is larger for
smaller $\beta$, whereas $F$ at low $\varepsilon$ is larger for larger $\beta$.
The range of $\varepsilon$ corresponding to these two different cases depends
on $r$, as can be deduced from the comparison between panel (a) and (b) in
Fig. 7. Since our sample objects are mainly distributed in the region of
higher $\varepsilon$ (solid histograms for the sample with radial velocities),
the probability is highest at smallest $\beta$.

\begin{figure}
\resizebox{\hsize}{!}{\includegraphics{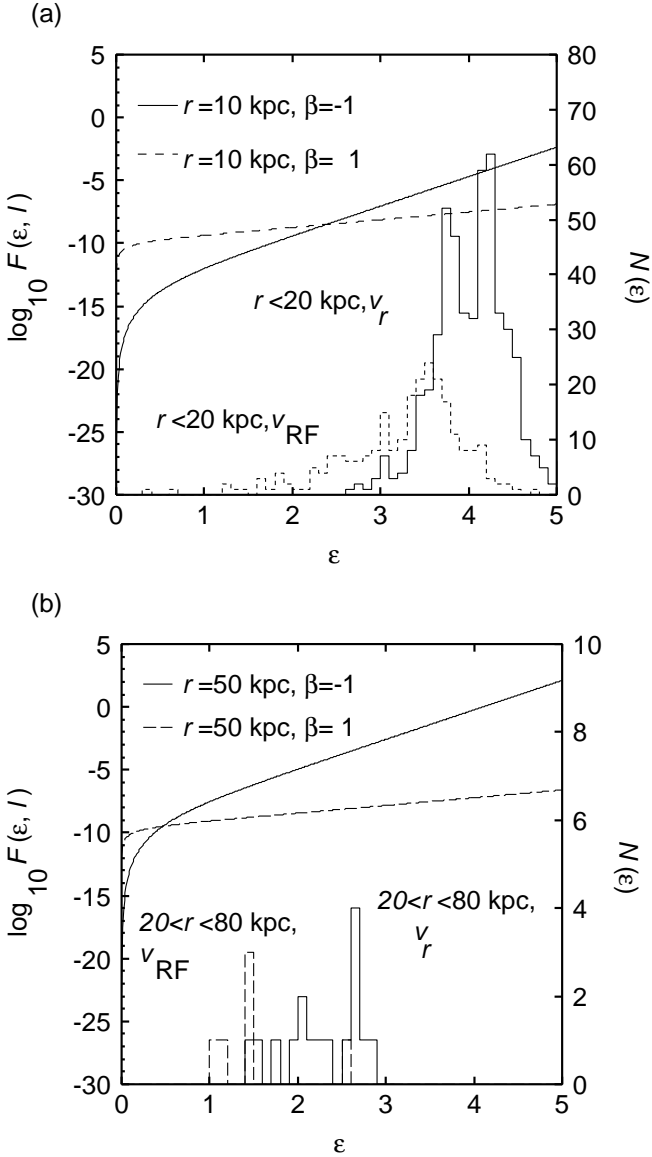}}
\caption{
The distribution function, $F$, for $\beta=-1$ (solid lines) and
$\beta=1$ (dotted lines), at $r=10$ kpc (panel a) and $r=50$ kpc (panel b).
Also plotted are the number distributions $N(\varepsilon)$ of the stars
when $a=200$ kpc, where dotted and solid histograms denote the sample with
and without available proper motions, respectively. The range of $r$ for
plotting $N(\varepsilon)$ ($r<20$ kpc for panel (a) and $20<r<80$ kpc for
panel b) is chosen to approximately match that for $F$.
}
\end{figure}

Following the above experiments, we drop the lower bound of $-1.5$ for $\beta$,
and search for the maximum probability at smaller $\beta$. No maximum is found
up to $\beta=-20$, although the large discrepancy in $M$ between the cases
with and without Leo~I remains. When we confine ourselves to the sample at
$r>10$ kpc, there exists a maximum probability at $\beta=-2.75$ (with Leo~I),
with a corresponding mass $32.0 \times 10^{11} M_\odot$.  For the
sample at $r>20$ kpc, we obtain $11.4 \times 10^{11} M_\odot$ at $\beta=0.8$.
This clearly suggests that the best-fitting $\beta$, obtained from
the analysis when only radial velocities are considered, is rather
sensitive to the range of $r$ employed in the sample selection.  This in turn
affects the number distribution, $N(\varepsilon)$, which is relevant to the
likely range of $F$ (Fig. 7).

With these unavoidable limitations of the present sample in mind, Table 2
summarizes the likelihood results for the limited range of $-1.5\le\beta\le 1$,
obtained for power-law and shadow tracers, using a variety of different priors
on $a$ and $\beta$.  The most likely value of $\beta$ is $-1.5$ for all cases,
for the reasons described above. We note that the current mass estimate is
rather {\it insensitive} to the $\beta$ prior. As the $\beta$ prior decreases,
the estimated mass generally increases, and the best-fitting $\beta$ decreases,
because the small $\beta$ prior is biased toward more tangentially anisotropic
velocity distributions than is the large $\beta$ prior. However, since most of
our sample have high $\varepsilon$, the best-fitting $\beta$ remains $-1.5$
{\it regardless} of whether we adopt the uniform prior or the uniform-energy
prior for $\beta$. This property makes the mass estimate insensitive to the
$\beta$ prior.

Now we apply the Bayesian likelihood method to the subsample of objects with
both radial velocities and proper motions available, and consider the derived
space motions. In contrast to the above case, where we used radial velocities
alone, we find that the maximum probability within the range of $\beta$
we consider is now {\it bounded} (Fig. 8a).  This may be caused by the
characteristic distribution of $\varepsilon$ for the sample with full space
motions, as shown in Fig. 7 (dotted histogram).  This figure shows that there
exists a larger fraction of low--$\varepsilon$ stars than are found in the
sample with radial velocities alone (solid histograms), so a larger $\beta$ is
preferred to achieve a larger $F$. The mass estimate in this case is quite
insensitive to the presence or absence of Leo~I.  Fig. 8b shows the
probabilities, as a function of $M$, with a fixed value of $\beta=-1.25$, for
the case of a power-law tracer population with $\gamma=3.4$.  Solid and dashed
lines denote the probabilities with and without Leo~I, respectively. As is
evident, the agreement between both probabilities is significantly improved
compared to the case when the radial velocities are considered alone (Fig.
6b).  When Leo~I is included, the most likely value of the total mass, $M$, and
the scale length, $a$, are $25.0\times 10^{11}M_\odot$ and 225 kpc, respectively.
Excluding Leo~I yields $M =18.0\times 10^{11}M_\odot$ and $a=160$ kpc.  Table 3
summarizes the various results obtained when the proper motions of the objects
are considered.  This Table illustrates that, for all cases, the mass of the
Galaxy obtained when including Leo~I is in good agreement with that obtained
without Leo~I.  Also, the mass estimate depends only weakly on the index
$\gamma$, unknown prior probabilities for $a$ and $\beta$, as well as on the
range of $r$ used in the sample selection, resulting in small changes in the
mass estimates over a range of only a few times $10^{11} M_\odot$.

\begin{figure}
\resizebox{\hsize}{!}{\includegraphics{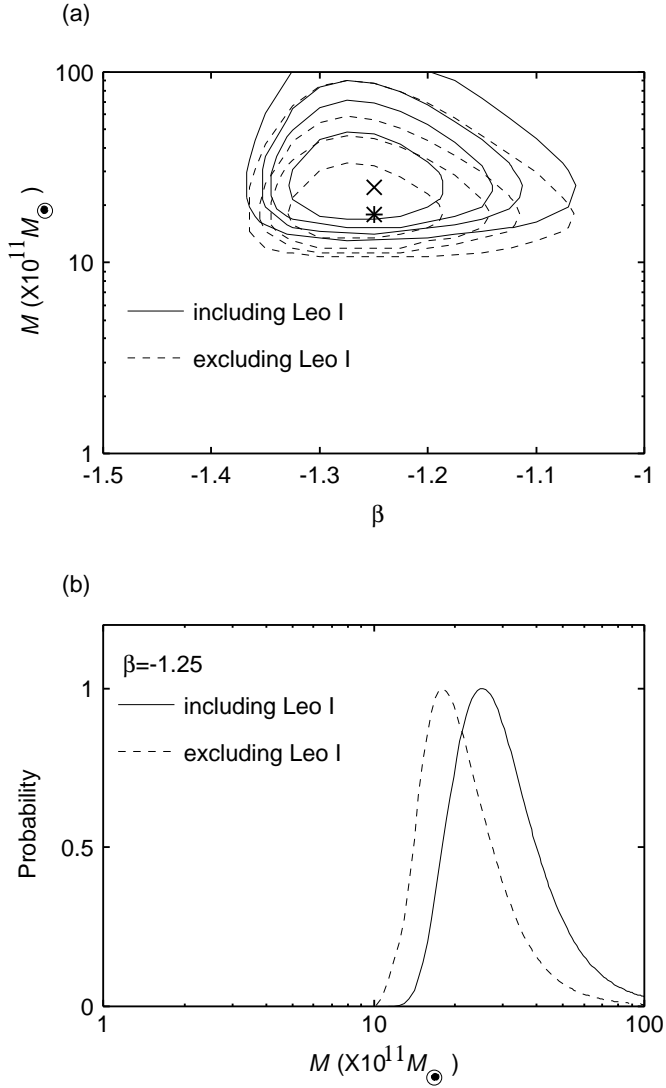}}
\caption{
(a) Likelihood contours in the plane of the mass, $M$, and velocity
anisotropy, $\beta$, obtained from an analysis that uses both radial velocities
and proper motions.  Solid and dashed curves show the results including Leo I
and excluding Leo I, respectively; the cross and the asterisk show the maxima
of the probabilities for each case. Contours are plotted at heights of 0.32,
0.1, 0.045, and 0.01 of the peak height. The spatial distribution of a tracer
population is assumed to follow a power-law form with $\gamma=3.4$.  (b)
Probabilities of the mass $M$ at the best-fitting $\beta$ of $-1.25$, including
Leo~I (solid line) and excluding Leo~I (dashed line).
}
\end{figure}

To estimate the typical errors in this mass determination that are associated
with the measurement errors of the 561 tracers we have analyzed, we have
conducted Monte Carlo simulations, adopting the assumptions that typical errors
in the distances and radial velocities are 10~\%, and 10 $\rm{km~s^{-1}}$,
respectively, and that the proper-motion errors are 1 $\rm{mas~yr^{-1}}$ for
globular clusters, 0.3 $\rm{mas~yr^{-1}}$ for satellite galaxies, and 5
$\rm{mas~yr^{-1}}$ for the FHB stars. We generated 561 data points (including
Leo~I) drawn from Gaussian distribution functions centered on the observational
data, and with dispersions set to the above typical errors. Given a true mass
$M$, or scale length $a$ (where we use $M=2.3\times 10^{12}~M_{\odot}$ with
$a=200$ kpc), and prior probabilities for $a$ and $\beta$ ($1/a^2$ and the 
uniform-energy prior, respectively), we calculate the most likely mass,
$M^\prime$, and compare it with an input true mass.  Fig. 9 shows the
distribution of the discrepancy between $M^\prime$ and $M$, $100 \times
(M^\prime - M ) / M$, obtained from 1000 realizations.  The error distribution
in the current mass estimate has a mean value shifted downward by $20$~\%, and
a dispersion of half-width 20~\%.  These values suggest that one might
adopt an estimate of the systematic error on the order of 20~\%, and a random
error of $\pm 20$~\%.  Exclusion of Leo~I does not influence the magnitude
of these errors.  It is worth noting that WE99 obtained roughly $\sim
100$~\% systematic errors, and $\sim 90$~\% random errors in their mass
estimate, which was based on about 30 data points.  The significant improvement
of our mass estimate is mainly due to our consideration of a much larger data
set that includes several hundred FHB stars.

\begin{figure}
\resizebox{\hsize}{!}{\includegraphics{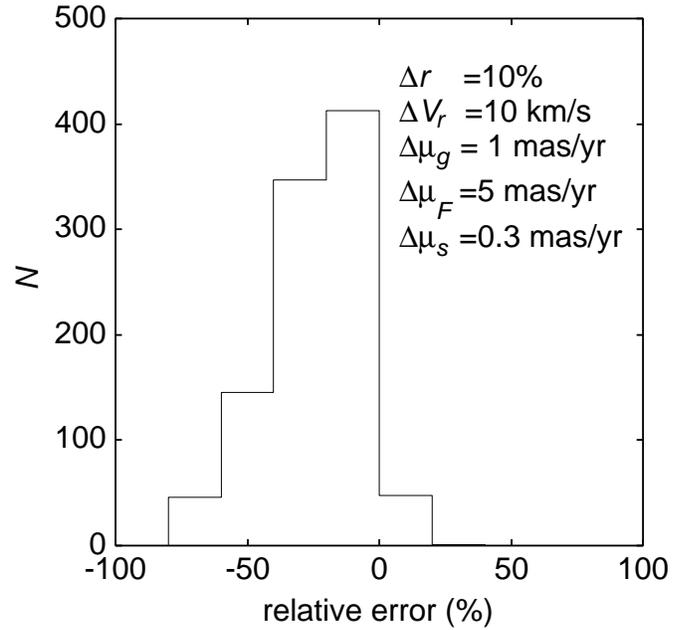}}
\caption{
An approximate error distribution of the mass estimate caused by
the typical measurement errors of the data. The abscissa denotes the relative
error in mass, $100\times(M^{\prime}-M)/M$, where $M^{\prime}$ is the mass
calculated by a Monte Carlo method and $M$ is the input true value. See
text for more details.
}
\end{figure}

As shown in Table 3, the most likely estimated total mass depends on model
assumptions at a level of a few times $10^{11} M_\odot$.  When the model is
fixed, the current large data set allows us to limit both systematic and random
errors to a level of about 20~\%. If we follow WE99's procedure for the
adoption of the most likely total mass, i.e., if we adopt the mass estimate
that provides the smallest difference between the masses obtained with Leo~I
and without Leo~I, we obtain $2.5^{+0.5}_{-1.0}\times 10^{12}~M_{\odot}$ (Leo~I
included) and $1.8^{+0.4}_{-0.7}\times 10^{12}~M_{\odot}$ (Leo~I excluded).  On
the other hand, the mass estimate within the distance of the LMC (50 kpc) is
quite robust, covering the narrow range 5.4 to 5.5 $\times 10^{11} M_\odot$.


\section{DISCUSSION AND CONCLUDING REMARKS}

We have placed new limits on the mass of the Galaxy, based on a newly 
assembled set of halo objects with the latest available proper-motion data.
First, the comparison of their space velocities with the escape velocities
at their estimated distances allowed us to show that the mass limits we
obtained depend on {\it neither} the presence or absence of Leo~I, {\it nor}
on the large velocity errors for Draco and Pal~3; a possible a lower limit on
the total mass of the Galaxy is about $\sim 2.2\times 10^{12} M_\odot$.
Secondly, a Bayesian likelihood approach has been used to derive a total mass
estimate for the Galaxy that is insensitive to the presence or absence of
Leo~I, at least when proper motions are taken into account. Although the best
mass estimate obtained from this approach depends somewhat on model assumptions
(prior probabilities for $a$ and $\beta$ and possibly the shape of $F$, see
below), the resultant systematic change of the total mass is confined to within
a few times $10^{11} M_\odot$.  The most likely total mass of the Galaxy we
derive is $2.5^{+0.5}_{-1.0}\times 10^{12}~M_{\odot}$.  This is in good
agreement with the total mass obtained by WE99 ($1.9^{+3.6}_{-1.7} \times
10^{12} M_\odot$) and that obtained from other methods  (e.g., Peebles 1995,
$2\times 10^{12}M_{\odot}$).  Since the size of our tracer sample is
significantly larger than used in previous studies, both systematic and random
errors are reduced to a great extent.  We note that consideration of the
numerous FHB stars plays a vital role in this mass estimate, as demonstrated in
Sect. 3.

It is also worth noting that, if we fix the mass of the Galaxy equal to our
most likely mass estimate, there is insufficient matter present to
gravitationally bind the LMC, if we adopt the recent proper-motion measurement
by Anguita et al. (2000).  These authors reported rather high proper motions,
$(\mu_{\alpha}{\rm cos}\delta, \mu_{\delta})=(+1.7\pm 0.2, +2.9\pm 0.2)$,
compared to previous measurements, $(\mu_{\alpha}{\rm cos}\delta,
\mu_{\delta})=(+1.94\pm 0.29, -0.14\pm 0.36)$ (Kroupa \& Bastian 1997).  Thus
their results need confirmation from other studies\footnote{Since we finished
our analysis here, a new study by Pedreros et al. (2002) has been
published [using the same method as Anguita el al. (2000)], that reaches
the same conclusion as Kroupa \& Bastian (1997).}.

The current work also implies that the Galactic rotation curve at outer radii,
$R > R_\odot$, does not decline out to at least $R \sim 20$ kpc (as long as
local disturbances to circular motions, such as warping motions and/or
non-axisymmetric motions, are ignored). As illustrated in Fig. 2, a declining
rotation curve corresponding to $a=20$ kpc and $V_{\rm LSR}=211$ km~s$^{-1}$ fails
to bind {\it many} sample objects to the Galaxy.  The smallest possible value
for $a$ to bind all objects in the isothermal-like density distribution (Eq. 4)
is $a=195$ kpc, yielding $V_{\rm LSR} \simeq 220$ km~s$^{-1}$.

In a more general context, the detailed shape of the rotation curve at and
beyond $R=R_\odot$ reflects the interplay between the disk and halo mass
distributions, as this region is located near the boundary of both components.
Thus, determining the rotation curve at $R_\odot \la R \la 15$ kpc
will set useful limits on the mass distribution in the inner parts of the
Galaxy. Indeed, the Japanese project VLBI Exploration of Radio Astrometry ({\it
VERA}) will be able to determine both inner and outer rotation curves from
the measurement of trigonometric parallaxes and proper motions of astronomical
maser sources that are widely distributed in the Galactic disk (Sasao 1996;
Honma et al. 2000).  {\it VERA} will reach unprecedented
astrometric precision, $\sim 10 \mu$as, and will yield precise determinations
of the Galactic constants $R_\odot$ and $V_{\rm LSR}$. We note that whatever
results are derived for the rotation curve, the total mass of the Galaxy ought
to be larger than $10^{12} M_\odot$, in order to bind the more distant stellar
objects.

Our estimate for the mass of the Galaxy inside 50 kpc, i.e., within the
distance of the LMC, is $5.5^{+0.0}_{-0.3}\times 10^{11}~M_{\odot}$ (Leo~I
included) and $5.3^{+0.1}_{-0.4}\times 10^{11}~M_{\odot}$ (Leo~I excluded).
The error estimates are calculated from the maximum and minimum values of the
total mass.  Thus, about 24\% of the total mass of the Galaxy resides within $r
\le 50$ kpc.  This implies that the possibility of brown dwarfs as the origin
of the microlensing events toward the LMC may be excluded, because it requires
a much smaller mass inside 50 kpc, $\sim 1.3\times 10^{11} M_{\odot}$ (Honma \&
Kan-ya 1998). Our result is also in good agreement with the recent statistics
of the microlensing events obtained from analysis of the 5.7-year baseline of
photometry for 11.9 million stars in the LMC (Alcock et al. 2000), showing the
absence of short-duration lensing events by brown dwarfs. However, the most
recent work has suggested that perhaps one of the microlensing events is
actually caused by a nearby low-mass star in the Galactic disk (Alcock et al.
2001). More direct observations for identifying lensing objects are required to
settle this issue.

Once the total mass of the Galaxy is fixed, it is possible to place a useful
constraint on the mass of the Local Group. Most of the mass in the Local Group
is concentrated in M31 and the Galaxy. The total mass of M31 can be estimated
from the positions and radial velocities of its satellite galaxies, globular
clusters, and planetary nebulae (Evans \& Wilkinson 2000; C\^{o}t\'{e} et al.
2000; Evans et al. 2000).  If we take it to be $1.2_{-0.6}^{+1.8}\times
10^{12}M_{\odot}$ (Evans \& Wilkinson 2000), the mass of the Local Group is
$\sim 3.7\times 10^{12}M_{\odot}$. This is in good agreement with the estimate
by Schmoldt \& Saha (1998), $(4-8)\times 10^{12}M_{\odot}$, based on modified
variational principles.

To set tighter limits on the total mass of the Galaxy we require more accurate
proper-motion measurements for a greater number of objects at large
Galactocentric distances. The high-velocity FHB stars in our sample (with
apparent magnitudes $V < 16$) that are responsible for setting the minimum mass
of the Galaxy have proper-motion errors of $\sim 5$ mas~yr$^{-1}$, whereas
Draco and Pal~3 have much larger {\it relative errors}, comparable to their
proper motions themselves (see Table 2).  Indeed, both the Space Interferometry
Mission ({\it SIM}: Unwin et al. 1997) and the Global Astrometry
Interferometer for Astrophysics ({\it GAIA}: Lindegren \& Perryman 1996) will
be able to provide more accurate proper motions for such high-velocity objects,
as well as for numerous other distant tracers of the Galaxy's mass, up to a
precision of a few $\mu$as for targets with $V \le 15$. This corresponds to an
error of $\la 10~{\rm km~s^{-1}}$ in the tangential velocity components
for many distant objects, i.e., comparable to the error of their (presently
determined) radial velocities.  Furthermore, roughly half of our sample objects
lack proper-motion measurements altogether.  To a great extent, the lack of
proper-motion measurements (at least for southern sources) will be removed with
the completion of the recently re-started Southern Proper Motion survey of
van~Altena and colleagues, as well as other efforts to substantially increase
the numbers of stars with reasonably well-measured proper motions (e.g., UCAC1:
Zacharias et al. 2000; UCAC2: Zacharias et al. 2001).

Further assembly of radial velocities for FHB stars, especially those at large
$r$ (beyond distances where accurate ground-based proper motions can be
obtained), is also of great importance for a number of reasons.  First, as
Fig. 3 demonstrates, large Galactocentric regions are characterized by small
escape velocities.  The current sample of FHB stars (because of
their locations near the Sun) explore distances where the corresponding escape
velocities are in the range of $500 \la V_{\rm esc}\la 600$ km~s$^{-1}$.
More remote FHB stars, with distances in the range $10 \la r \la 50$
kpc, will offer a further constraint on the total mass of the Galaxy by
covering the range $400 \la V_{\rm esc} \la 500$ km~s$^{-1}$. Secondly,
the assembly of samples of more distant FHB stars will enable exploration of
the suggested change in velocity anisotropy from the inner to the outer halo
(e.g., Sommer-Larsen et al.  1997), and better constrain its dependence on
Galactocentric distance.

In exploring the Bayesian approach for mass estimates of the Galaxy, we have
adopted a specific form of the phase-space distribution function $F$ (Eq. 8) to
facilitate comparison with previous studies.  This procedure implicitly assumes
that the velocity-anisotropy parameter, $\beta$, is constant everywhere in the
Galactic volume. However, as noted by Sommer-Larsen et al. (1997), there is an
indication that the velocity anisotropy of the halo may be mostly radial at $R
\la 20$ kpc and tangential at $R \ga 20$ kpc. If so, many of distant
FHB stars, especially those at $R > 20$ kpc, play a crucial role in the
determination of the global distribution of velocity anisotropy.  Searches for
a more realistic form of the phase-space distribution function, combined with a
more elaborate likelihood method, are both worthy pursuits.  
Also, instead of exploring such a specific but realistic form of
distribution function, a non-parametric method as proposed by Merritt \&
Tremblay (1993) will be more useful if a much larger data set is available.

Moreover, the implicit assumption behind the current method, that the sample
stars have random distributions in location and in their space motions, may not
be well satisfied if the halo is largely dominated by coherent structures such
as tidal streams (e.g., Ibata et al. 2001).  Alternative approaches to
obtaining mass limits using tidal streams (Johnston et al. 1999) are worth
considering in such a case.

Fortunately, prospects are excellent for obtaining a rapid increase in the
observational database of FHB stars with the required data.  There already
exists a substantial body of additional spectroscopy for FHB/A stars observed
during the course of the HK survey of Beers and colleagues and the Hamburg/ESO
Stellar survey (Christlieb et al. 2001), many of which also have available
proper motions, or will soon, from completion of the SPM survey and/or other
ground-based efforts.  However, as was noted by Wilhelm et al. (1999a)
(foreshadowed by Norris \& Hawkins 1991; Rodgers \& Roberts 1993, and
references therein; Kinman et al. 1994; Preston et al. 1994), a substantial 
fraction (perhaps as high as 50\%) of high-latitude A-type stars are 
{\it not} FHB, but rather some (as yet undetermined) mixture of binaries and 
high-gravity stars (see Preston \& Sneden 2000).  For some applications, such 
as estimates of the mass of the Galaxy that rely on space motions of tracers 
(and in turn on reasonably precise distance estimates of individual objects), 
confident separation of bona-fide members of the FHB population from 
possible ``contaminants'' is crucial
\footnote{For example, if 10~\% of our FHB sample is contaminated by
blue metal-poor stars, we obtain a $2 \sim 3 \times 10^{11} M_\odot$ decrease
in our total mass estimate, based on Monte Carlo experiments.}. In the past,
this has required that one obtain either Str\"omgren photometry and/or
spectrophotometry (e.g., Kinman et al. 1994), broad-band $UBV$ photometry in
combination with medium-resolution spectroscopy (e.g., Wilhelm et al. 1999a),
or reasonably high S/N, high-resolution spectroscopy (e.g., Preston \& Sneden 
2000).
All such endeavors are rather time intensive.  However, Christlieb et al.
(2002, priv. comm.) have been exploring means by which adequate separation
of FHB stars from higher-gravity A-type stars might be accomplished {\it
directly} from objective-prism spectra, such as those in the Hamburg/ESO
stellar survey.  Such methods, which look promising, would be most helpful in
future investigations of this sort.  Wide-field stellar surveys, such as those
presently being carried out with the 6dF facility at the UK Schmidt Telescope,
are capable of providing large numbers of radial velocities for FHB/A
candidates, and are expected to contribute 5,000-10,000 suitable data over the
course of the next few years.

\begin{acknowledgements}
We are grateful to B. Fuchs, R. B. Hanson, and I. Platais for assistance with
the comparison of the Wilhelm et al. sample with the catalogs of STARNET, NPM1,
and SPM~2.0, respectively.  We also thank the members of the VERA team for
several useful comments on this work.  T.C.B. acknowledges partial support from
grants AST-00 98508 and AST-00 98549 awarded by the U.S. National Science
Foundation.  T.C.B also would like to acknowledge the support and hospitality
shown him during a sabbatical visit to the National Astronomical Observatory of
Japan, funded in part by an international scholar award from the Japanese
Ministry of Education, Culture, Sports, Science, and Technology, during which
initial discussions of this work took place.
\end{acknowledgements}


\begin{table*}
\end{table*}
\begin{table*}
\includegraphics{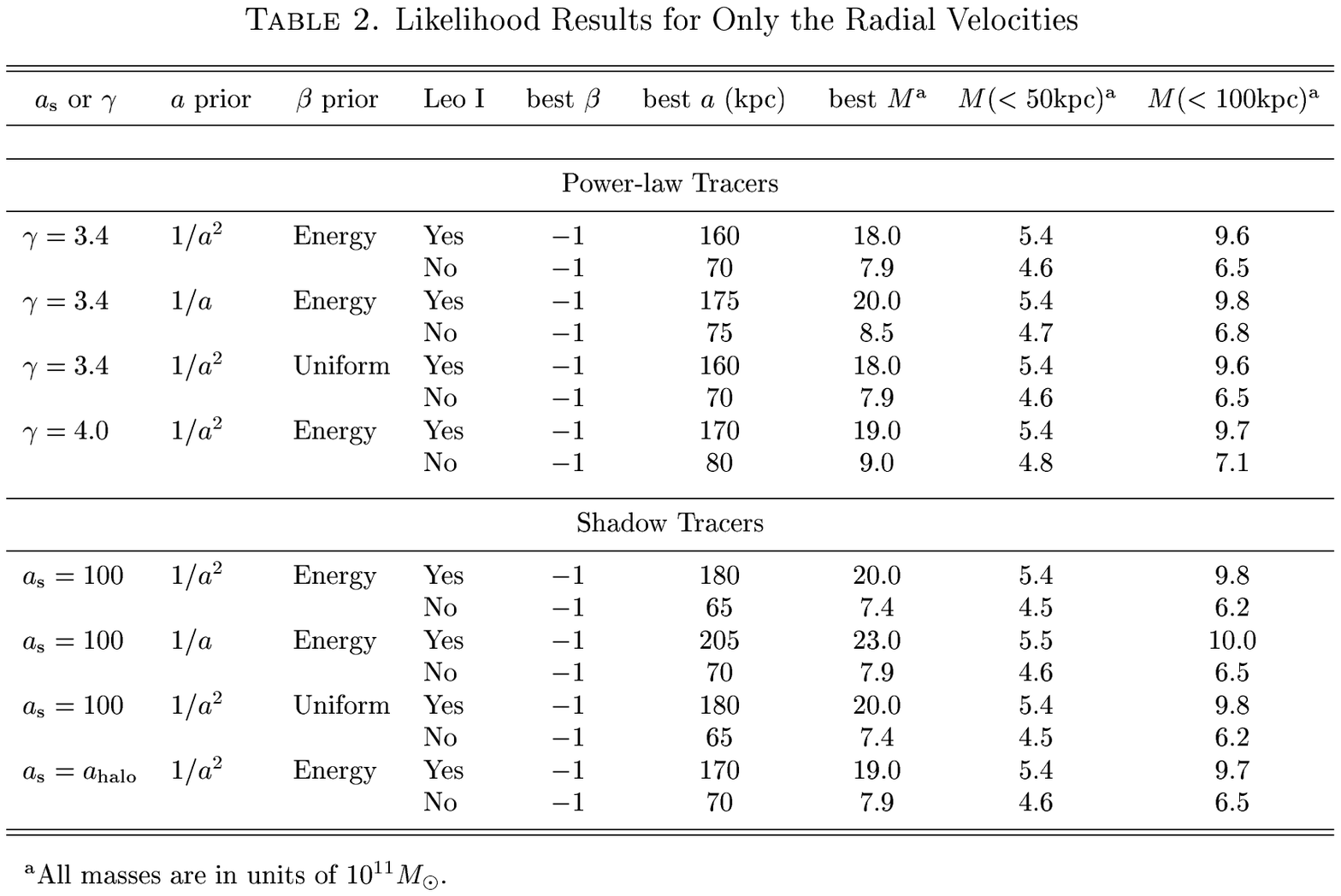}
\end{table*}
\begin{table*}
\includegraphics{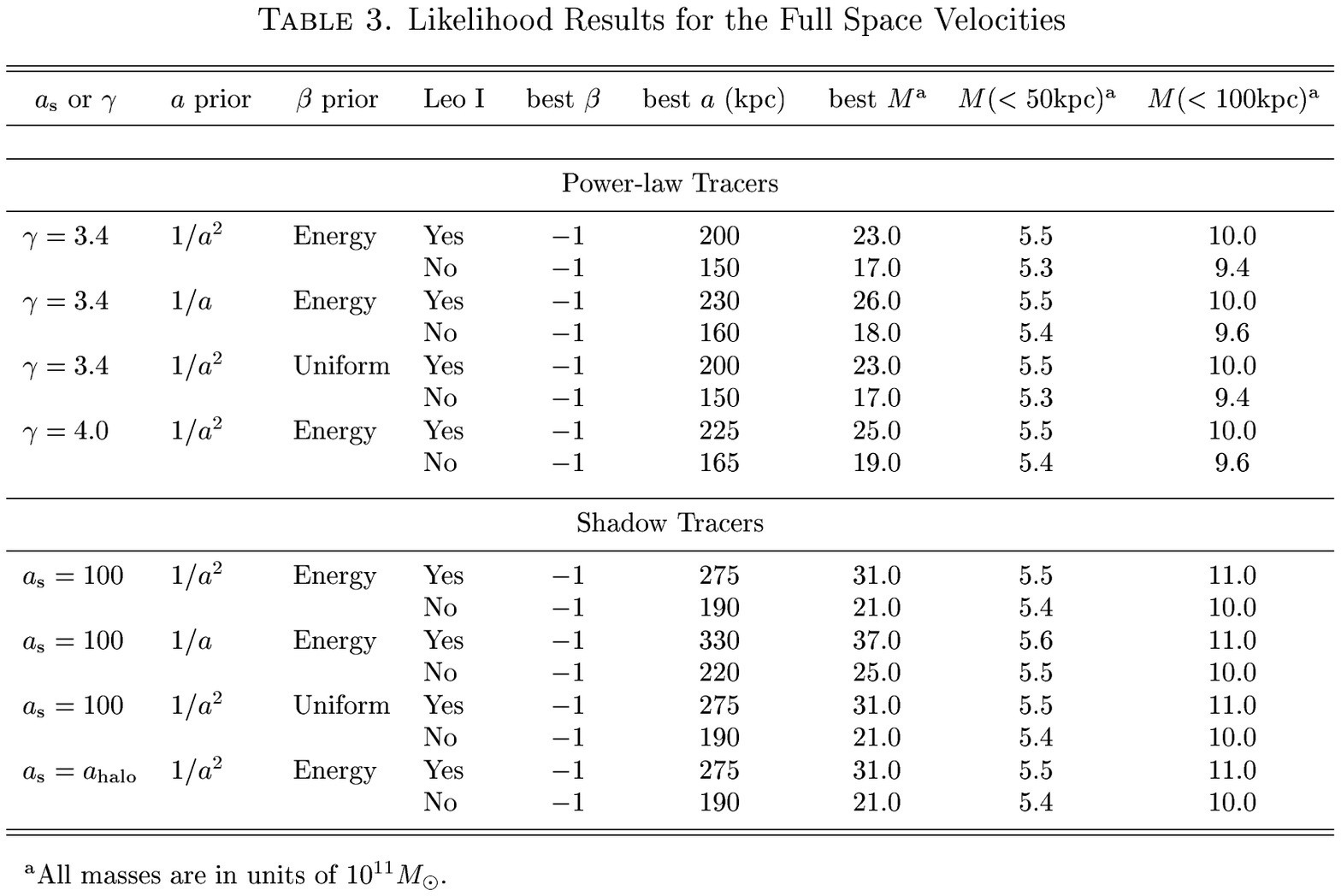}
\end{table*}




\end{document}